\documentclass[12pt]{iopart}
\eqnobysec
\usepackage{graphicx}
\begin{document}

\font\amsb=msbm10
\def\hbar{\mbox{\amsb\char'175}}
\def\hslash{\mbox{\amsb\char'176}}

\title[Decoherence of Wigner Functions]{Decoherence of Semiclassical Wigner 
Functions}

\author{Alfredo M. Ozorio de Almeida}

\address{ Max Planck Institute for Complex Systems, Noethnitzer Str. 38, 
D-01187 Dresden, Germany\footnote[1]{Gutzwiller Fellow}}

\address{Centro Brasileiro de Pesquisas Fisicas, Rua Xavier Sigaud 150,
22290-180, Rio de Janeiro, RJ, Brazil\footnote{permanent address}}

\begin{abstract}
The Lindblad equation governs general markovian evolution of the density operator
in an open quantum system.
An expression for the rate of change of the Wigner function as a sum of integrals
is one of the forms of the Weyl representation for this equation. 
The semiclassical description of the Wigner function in terms 
of chords, each with its classically defined amplitude and phase, is thus inserted in 
the integrals, which leads to an explicit differential equation for the
Wigner function. All the Lindblad operators are assumed to be represented by smooth phase
space functions corresponding to classical variables. In the case that these are real, 
representing hermitian operators, 
the semiclassical Lindblad equation can be integrated. There
results a simple extension of the unitary evolution of the semiclassical Wigner function,
which does not affect the phase of each chord contribution, while dampening its amplitude.
This decreases exponentially, as governed by the time integral of the square difference
of the Lindblad functions along the classical trajectories of
both tips of each chord. The decay of the amplitudes is shown to imply diffusion in
energy for initial states that are nearly pure. Projecting the Wigner function 
onto an orthogonal position or momentum basis, the dampening of long chords emerges 
as the exponential decay of off-diagonal elements of the density matrix.

\end{abstract}


\submitto{\JPA}

\maketitle

\section{Introduction}

{\it Wigner functions}, $W(x)$, represent a density operator, $\hat{\rho}$, 
in the phase space,
$x=(p,q)=(p_{1},..., p_{l}, q_{1},..., q_{l})$, for a system with $l$ 
degrees of freedom. Given the position representation
$\langle q_+| \hat{\rho} |q_-\rangle$, the {\it Weyl-Wigner transformation}
defines \cite{Wigner}
\begin{eqnarray}
W(x)\equiv \int{dq'\langle q+\case12q'|\frac{\hat{\rho}}{(2\pi\hbar)^l}|q-\case12q'\rangle
\>\exp \left (-{\rmi\over\hbar}p\cdot q' \right )}.  
\label{def}                  
\end{eqnarray}
If we replace within (\ref{def}) $\hat\rho/(2\pi\hbar)^l$ by an arbitrary 
operator $\hat B$ that acts on the Hilbert space of the quantum system , the
resulting function, $B(x)$, is then known as the {\it Weyl representation}, or the
{\it Weyl symbol} for $\hat B$.
 
The semiclassical approximation for pure states,
$\hat{\rho}=|\psi\rangle\langle\psi|$, has the familiar WKB form
\begin{eqnarray}
\langle q|\psi\rangle \approx \sum_j a_j(q)\> 
\exp \left [{\rmi\over\hbar} s_j(q) \right ], 
\label{state}                                                                    
\end{eqnarray}
where the $s_j$'s are classical actions defined on a classical manifold, 
i. e. the integral of $p\cdot \rmd q$ from some arbitrary position.                                                                                 
The corresponding Wigner function, obtained by inserting (\ref{def}) in \eref{state} 
and evaluating the integral by stationary phase, can be decomposed into
\begin{equation}
W(x)\approx\sum_j\widetilde{W}_k(x),
\label{decompure}
\end{equation}
where each term has the form
\begin{eqnarray}
\widetilde{W}_k(x)=A_k(x) \> \cos{S_k(x)\over\hbar}.
\label{SCWpure}
\end{eqnarray}
The sum in \eref{decompure} runs over all {\it chords} centred on $x$ that connect 
appropriate paths
on the classical manifold manifold that corresponds to $|\psi\rangle$. For instance, in
the case where $l=1$, this is just the Bohr-quantized energy shell and $S_k$ is 
the area between the chord and the shell (plus a semiclassically small {\it Maslov phase})
as shown in Figure 1.  Both semiclassical approximations \eref{state} and \eref{SCWpure}
break down along {\it caustics}, where the amplitudes $a_j$, or
$A_k$, display spurious divergences. The caustics of Wigner functions are the loci
of coalescing chords. It is there nescessary to resort to uniform approximations
for the integral (1.1), resulting in Wigner functions that have large but finite
values in the neighbourhood of the caustics. Detailed studies of these are given by
Berry \cite{Berry1}, for $l=1$, and  in \cite{OAH} for general integrable systems. 
The various expressions for the semiclassical amplitudes in \eref{SCWpure}, also given
by these references, are presented in the Appendix. The verification that
\eref{SCWpure} is consistent with the {\it pure state condition}, $\hat\rho^2=\hat\rho$, 
within the semiclassical approximation, irrespective of caustics, was
carried out in \cite{OA}.

\begin{figure}
\begin{center}
\includegraphics[width=11.0cm]{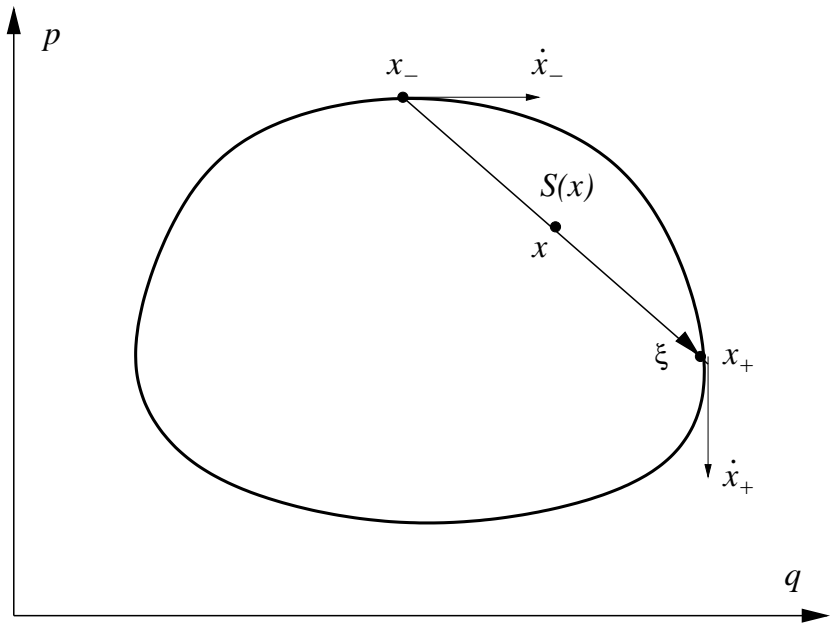}
\end{center}
\caption{A single chord, $\xi$, is centred on $x$ if it is close to a convex energy shell
in the case of one degree of freedom. The phase of the Wigner function is proportional
to the area $S(x)$ between the chord and the shell, while the amplitude, $A(x)$, depends
on both phase space velocities at the tips of $\xi$, as described in the Appendix. }
\label{fig1}
\end{figure}

In order to describe mixed states,
\begin{eqnarray}
\hat{\rho}=\sum_n c_n |\psi_n\rangle\langle\psi_n|,
\label{Wmixed}
\end{eqnarray}
we merely superpose the corresponding semiclassical Wigner functions. This is
simpler than in the position representation, in which we must multiply the sums 
for the bras and the kets. Indeed, there are cases where the semiclassical 
approximation to the mixed Wigner function is approached directly, instead of
relying on the pure states \cite{BalazsZ}. Such is the case of the {\it spectral Wigner 
function} \cite{Berry2}. This is defined as
\begin{eqnarray}
{\cal W}(x;E,\epsilon)\equiv\frac{(2\pi\hbar)^l}{\sqrt{2\pi}\>\epsilon}\>
\sum_n W_n(x)\> \exp\left[-\frac{(E-E_n)^2}{2\epsilon^2}\right],
\label{Wspectral}
\end{eqnarray}
where $W_n(x)$ is the pure state Wigner function for the n'th eigenstate of a
given Hamiltonian. If the width of the {\it energy window}, $\epsilon$, is classically
small, but contains many eigenstates, then the semiclassical limit of
\eref{Wspectral} can also be decomposed into
\begin{equation}
{\cal W}(x;E,\epsilon)\approx\sum_k\widetilde{{\cal W}}_k(x;E,\epsilon),
\label{decomspec}
\end{equation}
where
\begin{eqnarray}
\widetilde{{\cal W}}_k(x;E,\epsilon)={\cal A}_k 
\exp\left[-\frac{\epsilon^2\tau_k^2}{2\hbar^2}\right]
\cos\left[{{\cal S}_k(x)\over\hbar}\right].
\label{SCWspectral}
\end{eqnarray}
Here ${\cal S}_k(x)$ is again the area closed off by a chord centred on $x$, but both
its tips are now further constrained to lie on the same classical trajectory, within 
the energy shell of energy $E$. The time taken for the traversal of this 
trajectory segment is $\tau_k$ and the amplitudes ${\cal A}_k$ are given in references
\cite{Berry2,reports}. If an alternative Laurencian energy window is used, 
such as in \cite{Tosca}, then the chord contributions decay exponentially, rather than 
with a gaussian factor. 
It is important to note that \eref{SCWspectral} holds for any 
kind of classical motion, whether integrable, or chaotic to any degree, since it 
depends only on finite times of the order of $\hbar/\epsilon$. In the case of 
chaotic motion, the semiclassical limit of the individual $W_n(x)$ in
\eref{Wspectral} is not known, but we can still work with the semiclassical 
mixture \eref{SCWspectral}. 

Even initially pure density operators evolve into a mixture if the system is open.
Thus the Wigner function of a system that attains equilibrium with a heat bath
at the temperature $T$ is given by the canonical distribution
\begin{eqnarray}
W_T(x)=\frac{1}{Z(T)}\sum_n\exp\left(-\frac{E_n}{kT}\right)\> W_n(x),
\label{thermalW}
\end{eqnarray}
where $Z(T)$ is the partition function. The semiclassical limit of $W_T(x)$,
in the case of $l=1$ was obtained by  Korsch \cite{Korsch}.

How can we take the external environment into account in the evolution of an 
initially pure state? The {\it influence functional} approach to this problem 
is to construct a specific
(simplified) model of the environment and its interaction with the system, so as
to subsequently trace out the external variables \cite{FV,CL}. Alternatively, the theory of
semigroups starts from the mathematical restrictions that the resulting mixed
density matrix must satisfy. A remarkable simplification follows from the assumption  
that the environment reacts to the system sufficiently fast so as to loose all 
memory of prior states of the system, i.e., the evolution is assumed to be 
markovian. The density operator must then satisfy a linear differential master
equation of the following form:
\begin{eqnarray}
\frac{\rmd\hat\rho}{\rmd t}=-{\rmi\over\hbar}[\hat H,\hat\rho]
+{1\over\hbar}\sum_j(\hat L_j\hat\rho\,\hat L_j^\dag
-\case12\hat L_j^\dag\hat L_j\hat \rho -\case12\hat\rho\,\hat L_j^\dag\hat L_j).
\label{Lind}
\end{eqnarray}
This theorem represents a culmination of work on open systems\cite{K,D,GKS,L} 
and is known as the {\it Lindblad equation}.

The first term accounts for the unitary evolution of 
the closed system under its own internal Hamiltonian operator, $\hat H$. 
The {\it Lindblad operators}, $\hat L_j$ are arbitrary in principle, 
but physically they must be
related to the observables which couple the internal system to the environment.  
The simplest example is that of a system that is subjected to an 
external field, so that the interaction depends on its position, but where there is 
no net exchange of energy with the environment. In this case, we can assume that
the $\hat L_j=c_j\hat q_j$, i. e. the Lindblad operators are proportional to the 
components of the position operator. The dependence of the coupling constants on
$\hbar$ is already divided out in \eref{Lind} as discussed in \cite{Perc}.
Once the master equation has been constructed on the basis of physical assumptions
about the way that the system is affected by the external world, the dynamical
variables of the environment play no further explicit role in the internal 
evolution. It is important to distinguish the Hamiltonian appearing in \eref{Lind},
which drives the internal motion, 
from the one that defines the pure, or mixed initial states in previous equations. 
If they happen to coincide, then the density operator will be constant in 
the absence of interaction with the environment. 

The purpose of this paper is to analyze the evolution of Wigner functions that
have initially the semiclassical pure state form \eref{decompure}, or the spectral 
form \eref{decomspec},
under the influence of an arbitrary Hamiltonian and arbitrary Lindblad operators.
It goes without saying that alternative methods to the decomposition
of master equations for the density matrix in any orthogonal basis are much more
valuable than for the Schroedinger equation itself, since the number of states in 
the truncated basis needs to be squared. This is the reason for the popularity
of stochastic formalisms reviewed in \cite{Perc}, but the simple adaptation
that needs to be made for the inclusion of hermitian Lindblad operators in the 
semiclassical evolution allows us to work directly with the Wigner function and its projections.
 
The fact that both the Weyl-Wigner transformation \eref{def} and the Lindblad equation \eref{Lind}
are linear with respect to superpositions of density operators allows us to
deal separately with each chord contribution, $\widetilde{W}_k(x)$, or 
$\widetilde{{\cal W}}_k(x;E,\epsilon)$, as if it represented a {\it chord density} $\hat\rho_k$.
The only added assumption is that both $\hat H$ and the $\hat L_j$ are represented
by smooth phase space functions in the Weyl representation. This is the usual case
for an isolated system, where the observable $\hat H$ is given in phase space
by a function that is at least very close to its classical limit.
Though lindbladian dynamics are less familiar, the more commonly considered
examples \cite{Perc} couple the internal system to the environment through observables, 
represented by real phase space functions, or,
for instance, the annihilation operator of the harmonic oscillator, which is 
represented by a complex (but smooth) function.

No attempt will be made to analyse the accuracy of the approximate Markovian 
evolution in
a concrete physical situation. This is merely an extension of the usual procedure
followed in ordinary hamiltonian quantum mechanics: Though no system is completely 
isolated, it is worthwhile to build general quantum theories under the 
assumption of 
isolation, with the eventual option of later including external interactions
\footnote{For a recent discussion of non-markovian  stochastic treatment, see e. g \cite{Gasp}.}. 
Likewise, we can now analyse the semiclassical limit of lindbladian dynamics, 
which in some cases should be a preliminary step in the study of more
complex non-markovian 
interactions with the environment. Evidently, general lindbladian dynamics 
(semigroups) also encompass the ordinary (unitary group) evolution of closed systems.
It will be shown that the effect of the environment is easily accommodated 
within the semiclassical evolution in the case of hermitian Lindblad operators.
The analysis of the evolution of a measure of decoherence, the {\it linear entropy}
is part of the subject of a companion paper \cite{SOA}, though the  numerical
examples there use finite time step equivalents of lindbladian evolution.

It is important to establish distinctions and connections between this paper and previous 
work on the semiclassical limit of open systems. Perhaps this is
the first time that the theory is developed entirely in phase space. Although 
Caldeira and Leggett \cite{CL} already make some use of the Wigner function, 
it appears only in  an auxiliary role. This is also the case of Grossmann \cite
{Gross} who discusses the semiclassical limit of the influence functional 
approach. A remarkable paper by Strunz \cite{Strunz} comes much closer to the 
present work. 
A path integral is there developed, but, unlike usual influence functionals,
it is constructed for markovian Lindblad evolution. 
The propagator is defined in the position representation, using the Wigner function,
Its semiclassical limit is  derived, but
this does not describe immediately the propagation of the semiclassical Wigner
functions themselves. Indeed, this is a tricky problem even for pure Hamiltonian
evolution, that has been fully understood only recently \cite{ROA, Osb}. The 
important point is that Strunz defines a {\it decoherence influence functional}
for pairs of arbitrary paths in phase space that is here shown to account for the
dampening of the amplitude of each chord of the semiclassical Wigner function. The
same decaying factor is easily seen to affect the off-diagonal elements of the
semiclassical density matrix in the position representation.

Thus, the next section reviews the unitary evolution of semiclassical Wigner
functions and discusses the purity of the states.  This is followed in section 3 
by the derivation of the Lindblad equation for the Wigner function 
without the hamiltonian term. The equation is integrated in the case of
hermitian Lindblad operators. 
Section 4 combines the hamiltonian and lindbladian evolutions of both preceding sections
to produce the semiclassical description of decoherence of an evolving Wigner function 
over a finite time interval. Energy diffusion of a stationary Wigner function in an
open system is discussed in section 5. Section 6 shows how the inverse Weyl-Wigner
transformation leads to a semiclassical approximation in the position representation
that is again described by the decoherence distance 
functional. The delicate issue of normalization is treated in the Appendix.

\section{Review of unitary evolution}
\label{Uevol}
The {\it Moyal Formula} \cite{Moyal}
\begin{eqnarray}
\frac{\partial}{\partial t}W(x)\approx\{H(x),W(x)\}+\Or(\hbar^2),
\end{eqnarray}
\label{Moyal}
where \{,\} denotes the classical {\it Poisson Bracket}, suggests that the
semiclassical evolution follows the classical trajectory for the argument of the
Wigner function. This is indeed the exact result for quadratic Hamiltonians and also
holds approximately if $W(x)$ is a smooth function within the scale of $\hbar$. 
For pure semiclassical states such as \eref{SCWpure}, or for the mixture
\eref{SCWspectral} this is certainly not the case. However, it has been shown that
the evolution then depends only on the orbits of both tips of each chord centred
on $x$ \cite{ROA,Osb}. Since neither of the two previous treatments of this problem
relied  directly on the differential equation for $\hat\rho_t$, i.e., the unitary
part of \eref{Lind}, it is a useful exercise to rederive the hamiltonian evolution 
in this way, preliminary to the inclusion of the new terms in the master equation.

The main ingredient is the integral form for the product of two operators $\hat B_2
\hat B_1$ in the Weyl representation, due to Berezin \cite{Berezin},
\begin{eqnarray}
B_2B_1(x)=\left(\frac{1}{\pi\hbar}\right)^{2l}\int\rmd x_2\rmd x_1 \>
B_2(x_2) B_1(x_1)\>\exp\left[{\rmi\over\hbar}\Delta(x, x_1, x_2)\right],
\label{Wproduct}
\end{eqnarray}
which was subsequently extended to higher order products \cite{pathint, reports}.
Here, 
\begin{eqnarray}
\Delta(x, x_1, x_2)=2(x\wedge x_1+x_1\wedge x_2+x_2\wedge x),
\label{triang}
\end{eqnarray}
where the {\it skew product} $a\wedge b=(\bi{J} a)\cdot b$, with
\begin{eqnarray}
\bi{J}=\left(\begin{array}{c|c} 0 & -1\\ \hline 1 & 0 \end{array}\right),
\label{symplectic}
\end{eqnarray}
the matrix that exchanges  positions and momenta in Hamilton's equations:
\begin{eqnarray}
\frac{\rmd x}{\rmd t}= \bi{J}\frac{\partial H}{\partial x}.
\label{Ham}
\end{eqnarray}
Each term in \eref{triang} is the (symplectic) area of the parallelogram defined by
the corresponding vectors (i.e. just the area itself for $l=1$). Much more useful is
to consider $\Delta$ as the symplectic area of the unique triangle that has 
$(x, x_1, x_2)$ as its midpoints. This is the {\it circumscribed triangle}, rather
than the usual  {\it inscribed triangle} defined by its corners. 

The equation that governs unitary evolution of the Wigner function is thus
\begin{eqnarray}                                                                             
\frac{\partial}{\partial t}W(x)=
{2\over\hbar}\left(\frac{1}{\pi\hbar}\right)^{2l}
\int\rmd x_2\rmd x_1\> H(x_2)W(x_1)\>
\sin\left[{\Delta\over\hbar}(x,x_1,x_2)\right].
\label{Hevol}
\end{eqnarray}
The evolution of semiclassical Wigner functions can be decomposed into
separate chord contributions. Hence  we insert \eref{SCWpure},
or \eref{SCWspectral} into \eref{Hevol} and evaluate the integral within the 
stationary phase approximation 
\footnote{The following development will be made for the pure Wigner
function \eref{SCWpure}. This holds also for the spectral Wigner function \eref{SCWspectral}, 
by including the 
exponential factor in the amplitude.}. Expressing the sines and cosines as the sum of
exponentials, there will be four integrals for each chord of the Wigner function:
\begin{eqnarray}
\fl\frac{\partial}{\partial t}\widetilde{W}_k(x)\approx -\frac{\rmi}{2\hbar}
\left(\frac{1}{\pi\hbar}\right)^{2l}
\sum_{\pm}\int &\rmd x_1\rmd x_2\> H(x_2)A_k(x_1)\>\nonumber\\
&\left\{\exp\left[{\rmi\over\hbar}[\Delta\pm S_k(x_1)]\right]
-\exp\left[-{\rmi\over\hbar}[\Delta\pm S_k(x_1)]\right]\right\}.
\label{SCHevol}
\end{eqnarray}
The stationary phase points are now obtained from the rule that the derivative,
$\partial/\partial x$, of a symplectic area, bounded by a chord $\xi$, is just
$\bi{J}\xi$ \cite{reports}. Hence, from
\begin{eqnarray}
(\partial/\partial x_1)[\Delta\pm S_k(x_1)]=-2\bi{J}\>(x_2-x)\pm\bi{J}\>\xi_k (x_1)=0
\nonumber\\
(\partial/\partial x_2)[\Delta\pm S_k(x_2)]=2\bi{J}\>(x_1-x)=0,
\label{stphase}
\end{eqnarray}
it follows  that the circumscribed triangle, $\Delta$, collapses at the stationary
configuration for which $x_1=x$ and $\xi_k(x_1)=\pm2(x_2-x)$, i. e. the Hamiltonian must 
be evaluated at the stationary points, which coincide with the tips of the chord 
centred on $x$.  At these points, the hessian
determinant of the phase is just
\begin{eqnarray}
\det\left[\begin{array}{cc}\frac{\partial^2 S}{\partial x_1\partial x_2} & -2\bi{J}\\
2\bi{J} & 0 \end{array}\right]=2^{4l}
\label{det}
\end{eqnarray}
and the signature of the hessian matrix (the number of positive, minus the number of
negative eigenvalues) is zero. To obtain this result, note that it obviously holds
for $S=0$ and we can always connect the eigenvalues of \eref{det} continuously to
this particular case (avoiding caustics). Since the eigenvalues remain finite
along any path and their product is unity, the signature remains zero.

Combining all these ingredients, we obtain the stationary phase evaluation of
\eref{SCHevol} as
\begin{eqnarray}
\frac{\partial}{\partial t}\widetilde{W}_k(x)\approx{1\over\hbar}
\left[H\left(x+{\xi_k\over2}\right)-H\left(x-{\xi_k\over2}\right)\right]
\> A_k(x)\>\sin{S_k(x)\over\hbar},
\label{Hevol2}
\end{eqnarray}
which can be immediately compared with the straightforward differentiation of
\eref{SCWpure}. The dominant term is
\begin{eqnarray}
\frac{\partial}{\partial t}W(x)\approx-{1\over\hbar} \sum_k
\frac{\partial S_k}{\partial t}\> A_k(x)\>\sin{S_k(x)\over\hbar},
\label{Hevol3}
\end{eqnarray}
which agrees with \eref{Hevol2}, provided that  
\begin{eqnarray}
-\frac{\partial S_k}{\partial t}
=H\left(x+{\xi_k\over2}\right)-H\left(x-{\xi_k\over2}\right).
\label{HJeq}
\end{eqnarray} 
But this form of the {\it Hamilton-Jacobi equation}, derived explicitly in
\cite{Osb}, is just the derivative of the action in \cite{ROA}. It differs
from Marinov's Hamilton-Jacobi equation which depends on just a single tip of the
chord \cite{Marinov}. 

Thus, we find that the stationary phase evaluation captures the dominant term of the
unitary evolution of the Wigner function, while neglecting the variation in
the amplitude $A_k(x)$. This propagation depends on the classical flow, through the
orbits of both tips of each chord, $\xi_k(x)$. (Only if $H(x)$ is quadratic, will the
difference of the hamiltonian vector fields at the chord tips in \eref{Hevol2} coincide 
with ${\bi J}\partial H/\partial x$ at the centre.)
Of course, we can also evolve the entire classical manifold and then 
reconstruct the Wigner function at the centre of the evolving chords as described by
Berry and Balazs \cite{BB}, resulting in a relatively slow change of amplitude 
\cite{ROA,Osb} that is not captured by \eref{Hevol2}.

To conclude this section, we note that both pure and mixed states can undergo
unitary evolution, but that, in the former case, the purity of an initial state
is not destroyed, i. e. we must have $\hat\rho^2=\hat\rho$ for 
all time if this is true initially. In the Weyl representation, the product rule
\eref{Wproduct} leads to
\begin{eqnarray}
W(x)=\left({2\over{\pi\hbar}}\right)^l\int\rmd x_1\rmd x_2\>W(x_1)W(x_2)\>
\exp\left[{\rmi\over\hbar}\Delta(x,x_1,x_2)\right],
\label{Pcond}
\end{eqnarray}
which was shown to hold  for general integrable systems 
within the stationary phase approximation \cite{OA}. It is quite
remarkable that this result is  entirely impervious to the multiplicity of chords
or their coalescense at caustics, which are washed out by the integration.

An important consequence of the fact that pure state Wigner functions satisfy the 
pure state condition is that we can take the trace of both sides to obtain
\begin{eqnarray}
\tr\hat\rho=\int \rmd x\>W(x)=(2\pi\hbar)^l\int\rmd x\>[W(x)]^2.
\label{trPcon}
\end{eqnarray}
It happens that the integral involving the square of the Wigner function does indeed 
equal unity, within the stationary phase approximation, while the first integral
does not, though its semiclassical value is independent of Planck's constant. These
results were first obtained by Berry \cite{Berry1}, for $l=1$, and are extended to
general integrable systems in the Appendix. It may be said that the pure state
Wigner functions are {\it indirectly normalized}.

\section{Pure lindbladian evolution}  

The Weyl representation of the remaining terms of the master equation \eref{Lind}
requires the integral form for the product of three operators
$\hat B_3\hat B_2\hat B_1$. Sometimes, as in \cite{ROA}, it is necessary to use
the full generalization of \eref{Wproduct}, integrating over three phase space
variables with a phase that is a circumscribed quadrilateral \cite{reports}.
However, in this case it is possible to use the following shortcut also provided by
\cite{reports}:
\begin{eqnarray}
\fl B_3B_2B_1(x)=\left({1\over\pi\hbar}\right)^{2l}
\int\rmd x_2\rmd x_1\> & B_3(x_2-x_1+x)B_2(x_2)B_1(x_1)
\exp\left[{\rmi\over\hbar}\Delta(x,x_1,x_2)\right].
\label{W3product}
\end{eqnarray}
Thus, the lindbladian evolution for a single chord of the semiclassical 
Wigner function \eref{SCWpure} becomes
\begin{eqnarray}
\fl\frac{\partial}{\partial t}\widetilde{W}_k(x) \approx &\hspace{-7mm}
{1\over{2\hbar}}\left({1\over{\pi\hbar}}\right)^{2l}\sum_{j,\pm}\nonumber\\
&\hspace{-7mm}\left\{\int\rmd x_2\rmd x_1\>
L_j(x_2-x_1+x)A_k(x_2)L_j(x_1)^*
\exp\left[{\rmi\over\hbar}\left(\Delta(x,x_1,x_2)\pm
S_k(x_2)\right)\right]\right.\nonumber\\
&\hspace{-7mm}\left. -\Re\int\rmd x_2\rmd x_1\>L_j(x_2-x_1+x)^*L_j(x_2)A_k(x_1)
\exp\left[{\rmi\over\hbar}\left(\Delta(x,x_1,x_2)\pm S_k(x_1)\right)\right]\right\}
\nonumber\\
\left.\right.
\label{Levol}
\end{eqnarray}
if we neglect the effect of the Hamiltonian. Here, $\Re$ stands for the real part
 and we note that 
an adjoint operator $\hat{B}^\dag$ is represented by the complex conjugate function 
${B(x)}^*$. Presuming that the Lindblad operators are represented by smooth  
phase space functions, the stationary phase evaluation of \eref{Levol} proceeds
essentially as in the previous section. Collecting the terms, we obtain
\begin{eqnarray}  
\frac{\partial}{\partial t}\widetilde{W}_k(x)\approx &{1\over\hbar} A_k(x)\sum_j
\left\{\Re\left[L_j(x+{\xi_k\over2})L_j(x-{\xi_k\over2})^*
\exp\left({\rmi\over\hbar}S_k(x)\right)\right]\right.\nonumber\\
&\left.-\case12\left[|L_j(x+{\xi_k\over2})|^2+|L_j(x-{\xi_k\over2})|^2\right]
\cos{S_k(x)\over\hbar}
\right\}.
\label{Levol1}
\end{eqnarray}

From here on the discussion will focus on the case of hermitian Lindblad
operators, i.e. it will be assumed that the system is only coupled to the
environment by observables. Then the master equation for $\widetilde{W}_k(x)$ 
simplifies to
\begin{eqnarray}
\frac{\partial}{\partial t}\widetilde{W}_k(x)\approx-{1\over{2\hbar}} 
A_k(x)\cos{S_k(x)\over\hbar}\>
\sum_j\left|L_j(x+{\xi_k\over2})-L_j(x-{\xi_k\over2})\right|^2.
\label{Levol2}
\end{eqnarray}
In contrast to the case of Hamiltonian evolution, we see that hermitian Lindblad 
operators do not affect the phase of the semiclassical Wigner function, but only 
its smooth amplitude function. Indeed, if we assume that the chords, $\xi_k(x)$, 
and the actions, $S_k(x)$, remain constant at  each point, $x$, we can integrate the 
equations for the amplitudes,
\begin{eqnarray}
\frac{\partial}{\partial t}A_k(x)=-{1\over{2\hbar}}
\sum_j\left|L_j(x+{\xi_k\over2})-L_j(x-{\xi_k\over2})\right|^2\> A_k(x),
\label{difampl}
\end{eqnarray}
to obtain
\begin{eqnarray}
A_k(x,t)=\exp\left[-{t\over{2\hbar}}\sum_j\left|L_j(x+{\xi_k\over2})
-L_j(x-{\xi_k\over2})\right|^2\right]\>A_k(x,0).
\label{evolampl}
\end{eqnarray}
Thus, it is easily verified that the semiclassical Wigner functions \eref{SCWpure}, 
now taken with the time dependent amplitude \eref{evolampl}, i. e.
\begin{equation}
\widetilde{W}_k(x,t)= A_k(x,t)\> \cos\>[S_k(x)/\hbar],
\label{Levol3}
\end{equation}
satisfy the evolution equation \eref{Levol2} and, of course, an equivalent result holds
for each chord of the spectral Wigner function, $\widetilde{{\cal W}}_k(x;E,\epsilon)$.   

Evidently the extreme case of pure lindbladian evolution will be relevant only if the
coupling constants implicit in the $\hat L_j$ are large enough for the internal
motion of the system to be effectively frozen within the decay time for the
amplitudes. The Lindblad operator with the largest difference between the tips of a
chord dominates the decay and the corresponding time is inversely proportional to the
square of the coupling constant.

The simplest case is that of a single operator $\hat L$. Clearly the Wigner function
remains constant if the state is initially an eigenstate of $\hat L$, since the tips
of all chords then lie on the same manifold $L(x)=${\it constant}. This result holds
for all the cases discussed in the Introduction: For $l=1$, this manifold is just a
phase space curve. For a general integrable state, the chord tips lie on an
{\it l}-dimensional torus, within the $(2l-1)$-dimensional surface, on which $L(x)=$
{\it constant} \cite{OAH}.  The chords of a spectral Wigner function
\eref{SCWspectral} also lie on 
the single surface $H(x)=${\it constant}. Hence, if $H(x)=L(x)$, the Wigner function
is also invariant.

Generally, for $H(x)\neq L(x)$, it must be that $L(x_+)\neq L(x_-)$ for most chords,
$\xi=x_+-x_-\>$.   Evidently, short chords are more stable against decay 
due to arbitrary Lindblad operators, but, for specific operators, there exist long 
chords that will
also be selected for survival. For instance, suppose that $l=1$ and $L(x)=cq$, the
position. Consider a given centre, $x$, with a vertical chord, $\xi(x)$, as in Figure 2,
then the amplitude does not decay. In this case, the full Wigner function condenses 
onto
the classical curve {\it and} onto the line of midpoints of the vertical chords that
bisects the classical curve. However, we shall see in the following section  that
this strange effect only arises in the complete absence of the hamiltonian term in
the evolution equation: No matter how strong the coupling to the environment, the 
Hamiltonian will eventually rotate the chords that are then erased by the Lindblad 
operator. For a general position coupling to the environment, with $l>1$, i.e. 
$\hat L=(c_1 q_1,...,c_l q_l)$, the decay rate will depend on the {\it distance}:
$[c_1(q_{1+}-q_{1-})^2+...+c_l(q_{l+}-q_{l-})^2]^{1/2}$.

\begin{figure}
\begin{center}
\includegraphics[width=11.0cm]{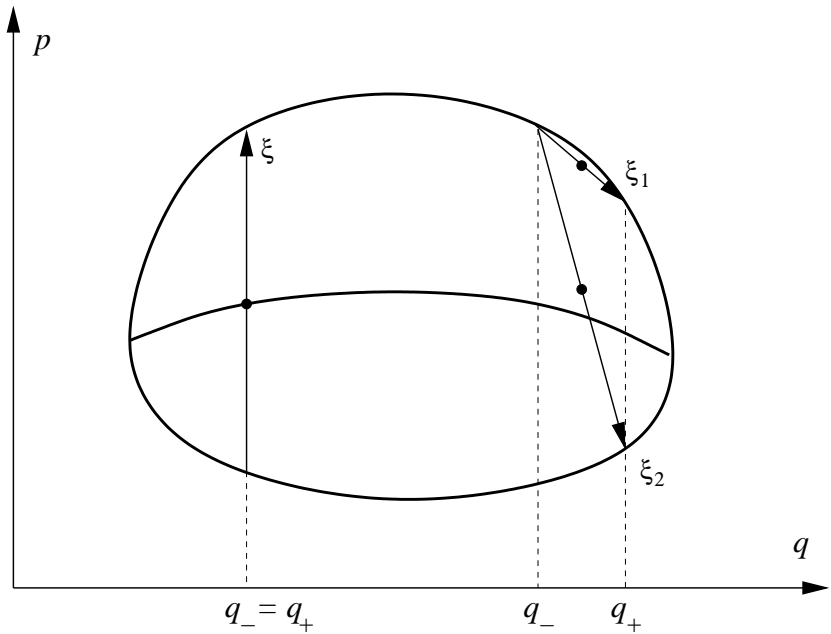}
\end{center}
\caption{Chords of the energy shell with the same $q_\pm$ can be short, $\xi_1$,
or long, $\xi_2$. Both will be equally dampened by the Lindblad operator $\hat{L}=cq$.
After a long time, the Wigner function condenses onto the energy shell {\it and} 
the line of midpoints of the vertical chords.}
\label{fig2}
\end{figure}

Evidently, the above picture is entirely consistent with previous results on
decoherence \cite{CL,Perc,Z}. The general initial pure state may be considered as a
superposition of the eigenstates of the Lindblad operator. The effect of 
"continuous measurement"  of this observable is to erode nondiagonal elements of
the density matrix in this representation \cite{Perc}. If we merely place a pair of
coherent states at both tips of a given chord, we obtain a Wigner function 
that is a gaussian at $x_{\pm}=x\pm\xi/2$ with a width of $\Or(\sqrt{\hbar})$.
The interference between these two states is represented by another gaussian centred
on $x$ itself but this is modulated by narrow fringes, of wavelength $\Or(\hbar)$.
This central gaussian decays exponentially just as the general semiclassical 
Wigner function that we have been studying when a Lindblad operator is 
"turned on", unless $L(x_+)=L(x_-)$. 

An initial semiclassical Wigner function  of the type \eref{SCWpure}, with the
amplitudes specified by \eref{ampl}, becomes mixed due the lindbladian decay of 
the amplitude
\eref{evolampl}. The way that this decoherence is connected with energy diffusion,
within the framework of the spectral Wigner function, 
will be discussed in Section 5. An important issue is the normalization of the 
mixed Wigner function that results from lindbladian evolution. It is shown in the
Appendix that the pure state itself is only indirectly normalized through
$\tr\hat\rho^2$, but this is now a measure of the decoherence of the mixed state
and should not equal $\tr\hat\rho$, once the environment is switched on. 
However, we find that the semiclassical evaluation
of $\tr\hat\rho$, though it is not unity, is stable with respect to the lindbladian
decay of amplitude, as well as with $\hbar$, taken as a small parameter. Therefore, we can
safely attribute the semiclassical change in $\tr\hat\rho^2$ as entirely due to decoherence.

\section{General evolution}
\label{Gevol}
We are now in a position to combine the unitary evolution of a closed quantum system
with the decoherence that results from its coupling to the environment, as modeled
by Lindblad operators.  Again we shall treat each chord contribution independently,
just as in the two previous sections. 
The differential equation corresponding to the master
equation \eref{Lind} is just the combination of \eref{Hevol2} with \eref{Levol2}
in the case of hermitian Lindblad operators, but it is hard to integrate this
directly. Instead, the evolution can be taken  as the result of alternate sequences
of unitary and pure lindbladian evolutions in the limit of arbitrarily small time
steps. This approach has been used for the coupling of two different unitary maps
by Berry et al. \cite{BBTV}. The recent study of a unitary map alternating with
a finite step lindbladian evolution by Bianucci et al. \cite{Bianucci} is even more
to the point. The important issue is that the exponential damping factor in the
amplitude remains constant during the unitary evolution described in section 2
through the classical motion of each chord and its centre. Alternatively, the chord is
then frozen during the interval in which the system is governed only by the Lindblad
operators.

Let us define $V_k(x,t)$ as the semiclassical contribution of each chord to the Wigner
function {\it in the absence of the Lindblad operators}, i. e. $V_k(x,t)$ describes
purely unitary evolution. Splitting the period of evolution, $t$, into $N$ half-steps
of hamiltonian evolution and $N$ half-steps of lindbladian evolution, we obtain
\begin{equation}
\fl\widetilde{W}_k(x,t)=V_k(x,t)\prod_{n=1}^N 
\exp\left[-\frac{t}{2\hbar N}\sum_j
\left|L_j\left( x_n+{\xi_k(x_n)\over 2}\right)-
L_j\left(x_n-{\xi_k(x_n)\over2}\right)\right|^2\right].
\label{gevol}
\end{equation}
Naturally, it has been necessary to multiply the Hamiltonian by a factor of $2$ 
in this formula, since it only acts during half of each interval, and the 
coupling constants in the Lindblad operators are likewise 
increased by a factor of $\sqrt2$. The final dampening of the amplitude at $x=x_N$
depends on both the history of the centres, $x_n$, and that of their chords, $\xi_k(x_n)$.
In other words, the total decay depends on the classical trajectories that arrive at
$x_{k,N\pm }=x_N\pm\xi_k(x_N)$. Each factor in the product is then calculated at the
centres, $x_{k,n}=(x_{k,n+}+x_{k,n-})/2$, with the chords, $\xi_{k,n}=x_{k,n+}-x_{k,n-}$,
during the half-intervals in which the hamiltonian is switched off. (Note that the
sequence of $x_{k,n}$'s depends on the final chord $\xi_{k}$, as well as on $x$.)
 
Taking the limit $N\rightarrow\infty$, we obtain the continuous evolution,
\begin{equation}
\widetilde{W}_k(x,t)=V_k(x,t)\> \exp\left[-{1\over{2\hbar}}\sum_j
\int_0^t\rmd t'\left|L_j(x_{k+}(t'))-L_j(x_{k-}(t'))\right|^2\right].
\label{gevol1}
\end{equation}
Therefore the full evolution of the Wigner function in an open system combines the
unitary evolution for each chord, obtained by the classical motion of the chord tips,
$x_{k+}(t)$ and $x_{k-}(t)$, with the exponential decay of the amplitude, characteristc 
of pure lindbladian decoherence. But now this dampening depends on the chord history.
If there is more than one chord centred on $x$,
each contribution to the evolution of the Wigner function depends on the
particular motion of both tips of each  respective chord. Generally the classical motion
of the centre of a chord does not coincide with the midpoint of the motions of the tips,
unless the motion is linear (i. e. the internal Hamiltonian is quadratic). Even though 
the development has focussed on the Wigner function of integrable systems, exactly the 
same form of general lindbladian evolution also holds 
for each chord contribution, $\widetilde{\cal W}_k$, 
to the spectral Wigner function \eref{SCWspectral}. The normalization of \eref{gevol1}
is discussed in the Appendix.

It is now feasible to make a partial connection to the path integral developed by
Strunz. Indeed, the {\it phase space influence functional} in \cite{Strunz} has exactly
the same form as the exponential in \eref{gevol1} in the case of hermitian Lindblad
operators, but this is there taken over all possible pairs of paths, which are not 
restricted to being classical trajectories. Even so, working within quite diverse 
frameworks, both treatments conclude that the effect of hermitian Lindblad operators 
is encapsuled by the {\it decoherence distance functional}:
\begin{equation}
D_t [x_+ (t'),x_-(t')]
\equiv\left[\sum_j\int_0^t\rmd t'\>|L_j(x_+(t')-L_j(x_-(t'))|^2\right]^\case12.
\label{functional}
\end{equation}
Though Strunz also discusses the semiclassical limit,
this is treated within the context of an {\it effective Hamiltonian}, 
$H(p,q)-{\rmi\over2}\sum|L_j|^2$, leading in general to complex trajectories. However,
in the specific example of the harmonic oscillator with the Lindblad operator 
$\hat L=c \hat q$, the classical paths in \cite{Strunz} that contribute to the 
propagator in the position representation turn out to be real.

The motion of the chords is important even for the Wigner function corresponding to an
eigenstate of the internal Hamiltonian, i. e. even if $V_k(x,t)$ is stationary. The tips
of each chord then evolve simultaneously along trajectories constrained to the invariant 
manifold (just the energy curve if $l=1$) that defines the semiclassical Wigner function. 
Thus, the dampening of the amplitude is only small for short chords, so that the line
of midpoints of vertical chords in Figure 2 is erased if the internal Hamiltonian
is taken into account. The same is true for spectral Wigner functions of the internal
Hamiltonian. Thus, unless all the $\{L_j,H\}=0$, only the classical region of the Wigner
function close to the energy shell survives the decoherence due to the external
environment. This general result is compatible with a previous specific deduction for
the harmonic oscillator \cite{ZHP}.

\section{Energy diffusion}

Let us now consider the ultimate effect of coupling a closed system to the environment
through Lindblad operators. The focus will be on an initially mixed state within a narrow
energy window, corresponding to a spectral Wigner function of the internal Hamiltonian. 
Of course, this also includes
a pure state, in the limit where the window width tends to zero, for the simplest case 
where $l=1$. The amplitudes of the various chord contributions to 
${\cal W}(x;E,\epsilon)$ in \eref{decomspec} decay exponentially with 
$(D_t [x_+(t'),x_-(t')])^2$ and the trajectories $x_\pm (t')$ are constrained to the
stationary energy shell. Hence, only short chords with centres very close to the 
shell will give appreciable contributions to this Wigner function after a finite interval
of time.

There are only two possibilities concerning the trajectories that connect the tips of 
a very small chord, $\xi_k$: Either this is the unique short trajectory that takes a 
small time,
$\tau_k$, between the two tips, or else the trajectory winds near a periodic orbit
\cite{Berry2,reports}. For $l=1$, the energy shell is itself a single periodic orbit,
so that the difference between both types of contributions concerns only the number
of repetitions around the periodic orbit (which is zero for the short orbit).
Both types of contribution will arise for all centres close to the shell in this simple
case. For $l>1$, the long orbits that contribute to the Wigner function with a small
$\xi_k(x)$ will be associated to neighbouring periodic orbits with varying actions, 
as $x$ is moved along the $(2l-1)$-dimensional shell. 
Thus, for any number of degrees of freedom, we can separate
\begin{equation}
{\cal W}(x,t;E,\epsilon)\mathop{\longrightarrow}_{ t\rightarrow\infty}
\widetilde{{\cal W}}_\tau (x,t;E,\epsilon)\>+\>\widetilde{{\cal W}}_{p.o.}(x,t;E,\epsilon),
\label{decomspec1}
\end{equation}
where we lump all the periodic orbit contributions into the second term. It is 
important to distinguish here the duration, $t$, of the evolution of the open system,
from the time, $\tau$, that the classical trajectory takes to travel between the tips 
of the short chord. Thus, $\tau$ is a function of $x$, for a fixed energy surface and is
completely independent of $t$.

It is easy to estimate the short orbit contribution, because
\begin{equation}
\xi_\tau (x)\approx\dot{x}\tau=\tau\bi J\>\partial H/\partial x,
\label{xi}
\end{equation}
to first order in $\tau$ \cite{reports}. Hence, we can approximate the difference of the
Lindblad functions in terms of a Poisson bracket:
\begin{equation}
|L_j(x+\xi_\tau /2)-L_j(x-\xi_\tau /2)|^2\approx\tau^2|\{H,L_j\}|^2.
\label{Lindif}
\end{equation}

Let us now assume that the closed classical system is ergodic. Then the time average
of $|\{H,L_j\}|^2$ for a trajectory on the energy shell exists 
and generally equals the average over the shell. 
But, for a short chord, the classical trajectory of the midpoint 
$x(t)\approx[x_+ (t)+x_- (t)]/2$, so that,
considering now $\tau$ to be an average time for the traversal from $x_-(t)$ to $x_+(t)$,
we can approximate the decoherence distance functional \eref{functional} so as to obtain,
\begin{equation}
\widetilde{{\cal W}}_\tau (x,t;E,\epsilon)\approx {\cal V}_\tau (x;E,\epsilon)\>
\exp\left[-\frac{t}{2\hbar}\sum_j\>\overline{|\{H,L_j\}|^2}\>\tau^2\right],
\label{Wspect}
\end{equation}
where we recall that in this case ${\cal V}_\tau$, the Wigner function for the closed system,
is stationary. Thus the short orbit contribution to the spectral Wigner function, after a
time $t$ of contact with the environment, has the same form as the  initial spectral Wigner
function \eref{SCWspectral}. However, the width of the energy window grows as
\begin{equation}
[\epsilon(t)]^2=\epsilon^2+ {\hbar t\over 2}\sum_j\>\overline{|\{H,L_j\}|^2},
\label{diffusion}
\end{equation}
i.e. the internal energy of the system spreads diffusively. This will also be the case for
an initially pure state if $l=1$. The absence of any net drift in the semiclassical
motion for a mixed state under the action of hermitian Lindblad operators is consistent
with previous work on the Lindblad master equation \cite{Perc}.

If $l>1$ a pure initial state will surely also
mix with neighbouring eigenstates of the other variables that together define its
invariant torus, but this effect is harder to retrieve. If the initial spectral Wigner
function is defined for a different observable, $\hat {\cal H}$, than the one 
that drives the internal motion of the system,
the surface ${\cal H}(x)=${\it constant}, which is the locus of the tips of all chords,
will no longer be stationary. In this case, we can no longer invoke ergodicity 
to estimate the growth
of the ${\cal H}_t$-energy window, though a similar argument can be used 
if the movement of the energy
shell is adiabatically slow as compared with the time that the orbits inside the shell require
to cover it. Of course, this approximately adiabatic scenario is more likely to arise
for $l=1$. In any case the decay of the
amplitude of the very short chords that survive the decoherence after a finite time will
still be ascribable to a net growth of the energy window.

If $\widetilde{{\cal W}_\tau}$ were the only term in the decomposition \eref{decomspec1} of the
spectral Wigner function, the effect of the Lindblad operators in the semiclassical
evolution would be reduced to merely mixing  the state in a widening energy window.
That is, the density matrix would be diagonal in the eigenbasis of the operator used in
the definition of the initial state. 
The periodic orbit terms in \eref{decomspec1} modify this picture in a way that requires
further analysis. In other words, {\it scars} of periodic orbits will play a significant
role in the semiclassical theory of decoherence, since their contribution to the evolved
Wigner function is not merely the same as in a spectral Wigner function that has 
a very narrow energy
window: Rather than being dampened by their period, the scars only decay as a funcion 
of the length of the chord, as measured by the decoherence distance functional.

\section{Position representation}

Instead of the basis of eigenstates of the internal Hamiltonian, let us consider
another orthogonal basis, such as the position representation. 
Then even the density matrix for an
initially pure state is viewed as a (coherent) superposition of a large number of basis
states. In the case of Lindblad operators that were just functions of the positions,
$\hat q$, it could have been easier to obtain the evolution of the semiclassical density matrix 
directly from the the position representation of the Lindblad equation \eref{Lind}.
For general  hermitian functions  of $\hat p$ and $\hat q$ it is better 
to proceed from the inverse Fourier transform to \eref{def}:
\begin{equation}
\langle q_+|\hat{\rho}|q_-\rangle=\int\rmd p\>\>W\left(p,\frac{q_+ +q_-}{2}\right)\>
\exp\left[{\rmi\over\hbar}p\cdot(q_+-q_-)\right].
\label{inverse}
\end{equation}

Sufficiently far from diagonality, i.e. for $q_+\neq q_-$, we need only be concerned with
the stationary phase evaluation in the oscillatory interior of the Wigner function,
so that the stationary phase condition is fulfilled where the chord centred at $x=(q_+ +q_-)/2$ 
has the component $\xi_q(x)=q_+-q_-$. The lindbladian dampening of the stationary chord can be factored out
of the integral, so that the density matrix of the evolved system is expressed in terms
of the decoherence distance functional \eref{functional} as
\begin{eqnarray}
\langle q_+|\hat{\rho}|q_-\rangle\approx\sum_{j_+,j_-}& a_{j+}(q_+)\>a_{j-}^*(q_-)\>
\exp\left[{\rmi\over\hbar}(s_{j+}(q_+)-s_{j-}(q_-))\right]\nonumber\\
&\exp\left\{-{1\over{2\hbar}}\{D_t[(q_+,p_{j+})(t'),(q_-,p_{j-})(t')]\}^2\right\}.
\label{qrho} 
\end{eqnarray}
Here the lable $j_{\pm}$ specifies the different momenta, $p_{j\pm}$, 
for the torus at the positions $(q_{\pm})$ and the actions
$s_{j\pm}$ are the usual actions in \eref{state}, measuring areas from the $q$-axes.

In the limit as $q_+\rightarrow q_+$, we also obtain
a short chord contribution from the classical region where the semiclassical approximation 
breaks down. However, the extrapolation of \eref{qrho} for short chords gives the same 
result as follows from the simpler approximation for the Wigner function \cite{Berry1},
$W(x)=(2\pi)^{-l}\delta(I(x)-{\cal I})$ for the eigenstate of an integrable Hamiltonian
corresponding to the action variables $I(x)={\cal I}$. Indeed, the correlation function
given in \cite{Berry3} follows from the short chord limit of \eref{qrho} by taking
$(s_{j+}-s_{j-})\rightarrow p_j\cdot(q_+ -q_-)$ as $q_+\rightarrow q_-$. Of course,
the decoherence distance functional vanishes in this limit, so that this contribution 
becomes insensitive to the Lindblad operators. We should note that there will generally also
be long vertical chords, such as shown in Figure 2, contributing to the density matrix. 
The projection of the semiclassical Wigner function onto position space is a good example of the 
nice behaviour of this approximation within integrals, despite the presence
of caustics.

The contribution of long chords to the density matrix of a mixture of energy eigenstates 
in a classically narrow energy window, corresponding to the spectral Wigner function,
has exactly the same form as \eref{qrho}. The decoherence distance functional dampens
the contributions from the various trajectories that satisfy the variational principle
for fixed energy between $q_-$ and $q_+$, so that, if we lable these by their initial and 
final momenta, $p_{j\pm}$, the stationary action is just $s_{j+}-s_{j-}$. The geometry
of the orbits and the semiclassical amplitudes of their contributions are thoroughly
discussed by Toscano and Lewenkopf in a recent paper \cite{Tosca}. They also provide the
limit for the contribution of short chords as $q_+\rightarrow q_-$, which tends 
to Berry's result \cite{Berry3} for $l>1$
\begin{equation}
{\cal W}(x)\propto\frac{J_{l/2-1}[p(q)|q_+ -q_-|/\hbar]}{[p(q)|q_+ -q_-|/\hbar]^{l/2-1}},
\label{correlation}
\end{equation}
where $J_\nu$ is a Bessel function. This result
follows from the approximation of the Wigner function as a delta function over the 
energy shell. Of course, there is no decoherence dampening for the null-chord contribution.

The semiclassical contributions from long chords to $\langle p_+|\hat\rho|p_-\rangle$ are
obtained by evaluating a similar integral to \eref{inverse} (see e.g. \cite{reports})
by stationary phase, with analogous results to \eref{qrho}. However, the limiting form 
for the contributions of small chords to the projection of the spectral Wigner function
does not generally have the same form as \eref{correlation}, because this was obtained
under the assumption of a simple $p^2$ dependence of the Hamiltonian.

The same general picture holds for any basis resulting from a linear canonical 
transformation, $x\rightarrow x'$, of the classical phase space, because of the well known
symplectic invariance of the Weyl-Wigner formalism (see e.g. \cite{reports}). 
The density matrix in the $q'$-representation follows from  
the transformation \eref{inverse}, after performing the classical change of variables. 
For each of these representations the decay of the off-diagonal elements
of the density matrix $\langle q'_+|\hat{\rho}|q'_-\rangle$ is determined by the
decoherence distance functional \eref{functional} evaluated at the tips of the
chords that project down precisely onto $q'_+$ and $q'_-$. This delicate quantum effect
quickly drives the density towards a diagonal mixture of the same initial set of basis 
states which were superposed coherently to form the initial state, if this was an
eigenstate of the internal Hamiltonian, because of the absence of drift for hermitian
Lindblad operators. The ultimate effect of the classical-like energy diffusion considered
in the previous section will be manifest only on a larger time scale. Again, it should
be pointed out that this qualitative scenario is well konwn (see e.g. \cite{CL,Perc,Z}),
the novelty being the complete model-independence in which these general semiclassical 
results are here obtained.

\section{Conclusion}

The evolution of quantum systems that are weakly affected by the external environment
can be studied within the framework of the Lindblad master equation \eref{Lind}. 
If the Lindblad operators
are all hermitian, the classical motion that supports semiclassical evolution  is
unaltered by an external environment, the only effect being to dampen the amplitude of
each chord contribution to the semiclassical Wigner function. In the present context,
this result embodies the known fact that hermitian Lindblad operators produce no net
drift \cite{Perc}. The decay of the semiclassical amplitudes was interpreted in 
section 5 as a diffusion of the energy of the initial (nearly pure) state, which
is a more obvious effect of opening the system than the delicate loss of quantum 
coherence that immediately manifests itself as a loss of amplitude in the highly 
oscilatory region of the Wigner function. An intuitive explanation for this decay
was recently put forward by Zurek \cite{Z1} in the restricted framework of interference
fringes in the Wigner function for a superposition of coherent states. 
These narrow oscillations arise near
the centre of the chord separating a pair of gaussians in phase space that would 
represent each single coherent state. Averaging over the juggling of the coherent states,
effected by the external environment, destroys the interfernce pattern. This same 
intuitive interpretation can now be generalized to the slight motion of the whole classical 
manifold that supports extended semiclassical quantum states.

General Lindblad operators also account for energy loss. This is in principle contained
in the differential equation \eref{Levol1} for the Wigner function, though no immediate 
separation of phase and amplitude is then possible. The path integral of Strunz 
\cite{Strunz} does lead to a semiclassical approximation in terms of complex trajectories.
However, this is derived in the position representation and even if its Wigner-Weyl transformation
\eref{def} were obtained, it would still not be a clear route to the semiclassical evolution of 
Wigner functions. The reason is that, even for unitary evolution, the propagation of the
Wigner function depends on the initial chord, through the orbit of both its tips. Thus,
the semiclassical propagator of Wigner functions is paradoxically irrelevant to their 
motion, as discussed in \cite{ROA}. In short, the general semiclassical description 
of the evolution of Wigner functions is undoubtedly more involved than the present 
restriction to hermitian Lindblad operators, i. e. observables. 

It is only in this case that the decoherence
distance functional \eref{functional}, defined by Strunz for arbitrary phase space paths,
localizes on the real classical trajectories with end points at the tips of each chord 
of the Wigner function. The simplicity of the Weyl representation of observables as
real smooth phase space functions then leads by direct integration to simple expressions
for the density matrix in the position, or the momentum representation, 
in terms of the same decay factor that accounts for the decoherence of Wigner functions.
The decay of large chords then appears as the dampening of off-diagonal elements 
in any representation.

Of course, one should be wary of the extrapolations that are required to synthesize
such a simple picture. The description of a spectral Wigner function corresponding to a
mixture of states in an energy window is only supported semiclassically   
by a single energy shell if the width of the window is classically small. Eventually,
energy diffusion will widen  the window beyond the limit of validity of this assumption.
But, anyhow, the basic hypothesis in deriving the whole semiclassical theory, that
only the oscillatory complex exponential terms in the integrals \eref{SCHevol} 
and \eref{Levol} were rapidly varying in phase space, eventually breaks down.
Indeed, for large evolution times the amplitude functions were found to become 
sharply peaked along the classical manifold that supports the semiclassical state.
Finally, one can add the further criticism that the simple semiclassical approximation,
based on independent chords, breaks down in the neighbourhood of the classical manifold 
on to which the Lindblad operators concentrate the Wigner function. There, the pure
states have been shown to be described by uniform approximations \cite{Berry1,OAH}
and ultimately their decoherent evolution should be ascertained.

Even so, the simplicity of a broad unified view over a wide range of diverse phenomena
is a rare enough asset to be thrown away easily! For instance, one is 
familiar with the many uses of the {\it Gutzwiller trace formula} \cite{Gutz} within various
branches of "semiclassics", in spite of the many conceptual problems with
its implementation. In the case of quantum spectra, it is also possible to find
improved approximations that smooth over spurious singularities ( see e.g. \cite{livro}),
but at a cost.The indirect
normalization of the square of the pure state Wigner function carried out in the Appendix
illustrates the uncanny selfconsistency of the simple semiclassical approximation,
which previously passed the more stringent test of the full pure state condition \cite{OA}.
The difficulties with chord multiplicities and caustics simply vanish within such integrals.
The companion paper \cite{SOA} shows, among other things, how the simple
semiclassical Wigner function here developed can be used to calculate the growth of
linear entropy, an important measure of decoherence. Indeed, the stability of the
direct normalization integral (also in the Appendix) throughout the action 
of the Lindblad operators, allows us to single out the decay of  
the Wigner function amplitudes as a pure decoherence effect resulting from the contact
of the internal system with the external environment.

\ack
I thank M Saraceno for  helpful discussions, as well as for originally directing
my attention to the lindbladian picture of open systems.
C  Rodrigues Neto and A N Salgueiro provided invaluable editorial assistance. 
Partial financial assistance from CNPq, Pronex and The Millenium Institute for 
Quantum Information is gratefully acknowledged.

\appendix
\section{Normalization}

The normalization of pure state semiclassical Wigner functions is a delicate matter.
Even though Berry \cite{Berry1} treated the case of one degree of freedom, it is 
worthwhile to rederive the integrals in a manner that is also valid for tori of higher
dimensions. The main idea, already employed in \cite{OA}, is to transform the integration
variables from centre coordinates, $x$, to the canonical angles $(\theta_-,\theta_+)$
at the tip of each chord on the quantized torus that is specified by its constant
action variables, $I_\pm={\cal I}$.

Let the points near each chord tip, $x_\pm$, take on local coordinates $X_\pm$, i. e.,
phase space points are given locally by $x=x_\pm +X_\pm$. Then one can also define
local action and angle variables as $I_\pm(X_\pm)\equiv I(x)$ and 
$\theta_\pm(X_\pm)\equiv\theta(x)$, so that near each chord tip,
\begin{equation}
\delta X_\pm=\frac{\partial X_\pm}{\partial\theta_\pm} \>\delta\theta_\pm
+\frac{\partial X_\pm}{\partial I_\pm}\>\delta I_\pm.
\label{dif1}
\end{equation}
Hence, for chord tips constrained to $I_\pm={\cal I}$, the Jacobian 
for the change of variables, $(\theta_-,\theta_+)\rightarrow \delta x=(X_+ +X_-)/2$,
is obtained from
\begin{equation}
\delta x=\case12\left.\frac{\partial X_+}{\partial\theta_+}\right|_{I_+=I}\delta\theta_+
        +\case12\left.\frac{\partial X_-}{\partial\theta_-}\right|_{I_-=I}\delta\theta_-,
\label{dif2}
\end{equation}
so that,
\begin{eqnarray}
\det\frac{\partial (\>p\>,\>q\>)}{\partial(\theta_-,\theta_+)}=
\det\left[\begin{array}{cc} {1\over 2}\frac{\partial p_+}{\partial\theta_+} &
{1\over 2}\frac{\partial p_-}{\partial\theta_-}\\
{1\over 2}\frac{\partial q_+}{\partial\theta_+} & 
{1\over 2}\frac{\partial q_-}{\partial\theta_-} \end{array}\right].                                   
\label{Jacobian}
\end{eqnarray}
Each element in \eref{Jacobian} represents an $(l\times l)$-dimensional block matrix and each
of its elements may be interpreted as a component, $p_1,...,p_l$, or $q_1,...,q_l$,
of the velocity vector $\dot x_\pm$ for the various choices of actions 
$I_1(x),...,I_l(x)$  as the Hamiltonian. In the case that $l=1$, these are proportinal
to the hamiltonian velocity vectors shown in Figure 1. Thus the Jacobian can be rewritten 
in the form
\begin{equation}
\left|\det\frac{\partial(\>p\>,\>q\>)}{\partial(\theta_-,\theta_+)}\right|=
\left({1\over 2}\right)^{2l}\>|\det\dot x_+\wedge\dot x_-|=
\left({1\over 2}\right)^{2l}\>|\det\{I_-,I_+\}|,
\label{Jacobian1}
\end{equation}
which is obvious in the case that $l=1$. In the general case this is obtained by the same
manipulations used in \cite{OAH} to prove the complementary relation:
\begin{equation}
|\det\{I_-,I_+\}|=\left|\det\frac{\partial(I_-,I_+)}{\partial(p_-,q_-)}\right|.
\label{Jacobian2}
\end{equation}
It is important to note that the matrix of Poisson brackets, $\{I_-,I_+\}$,
is only $l\times l$, whereas the Jacobians in both \eref{Jacobian1} and \eref{Jacobian2} 
have $(2l)\times(2l)$ dimensions.  
   
There is just one chord for each
pair $(\theta_-,\theta_+)$, but the same chord results from the interchange of 
$\theta_-$ with $\theta_+$, so that the overall transformation of the integration 
variables is
\begin{equation}
\int\>\cdot\>\>\rm dx=2^{-(2l+1)}\int_0^{2\pi}\cdot\>\>\>
|\det\{I_-,I_+\}|\>\rm d\theta_-\rm d\theta_+.
\label{trintegral}  
\end{equation}  
In passing from integrals over centres to integrals over pairs of angles, 
all problems with chord multiplicities and even caustics disappear. 
The chord densities really take on an independent existence, so that from this new
point of view, they may merely happen to share the same centre. Whereas in the
original semiclassical Wigner function a pair of chords with the same centre coalesce
as the centre moves on to a caustic, here the amplitude for the single chord determined
by $\theta_-$ and $\theta_+$ becomes singular but the integral is regularized 
by the determinant in \eref{trintegral}. 

It is easier to deal first with the integral over the square of the Wigner function
in \eref{trPcon}. Introducing in \eref{SCWpure} the amplitude of a chord contribution 
to the semiclassical pure state Wigner function as \cite{OAH}
\begin{equation}
A(x)= {2\over{(\pi\sqrt{2\pi\hbar})^l}}\>|\det\{I_-,I_+\}|^{-\case12},
\label{ampl}
\end{equation}
and averaging over oscillations, i. e. taking $\cos^2=\case12$, we immediately obtain
\begin{equation}
\tr\hat\rho^2=\int\frac{\rm d\theta_-}{(2\pi)^l}\>\frac{\rm d\theta_+}{(2\pi)^l}=1.
\label{rhosq}
\end{equation}

Given that the pure state condition is verified semiclassically, \eref{rhosq}
indirectly should imply that, also $\tr\hat\rho=1$. However, this is not the result 
of the direct calculation of
\begin{equation}
\fl\tr\hat\rho=\int\rm dx\> W(x)\>\approx\>\frac{1}{(4\pi\sqrt{2\pi\hbar})^{\it l}}\>
\int\rm d\theta_-\rm d\theta_+\>|\det\{I_-,I_+\}|^\case12\>
\cos\left[\frac{S(\theta_-,\theta_+)}{\hbar}\right].
\label{trrho}
\end{equation}  
This integral is dominated by the region where $\theta_+$ is near $\theta_-$, even
though the determinant in the r.h.s. of \eref{trrho} cancels for $\theta_+=\theta_-$, 
because there the hessian determinant of $S(\theta_-,\theta_+)$ also tends to zero.
It is first necessary to estimate the area, $S$, for small chords. Let us return to the
interpretation of a given angle, $\theta_\nu$, as the time along a trajectory on the 
torus, generated by choosing the respective action variable, $I_\nu$, as the Hamiltonian. 
If we transfer the origin to the torus, at $\theta_-$, so that $p_-=q_-=0$, 
the action for the chord to $\theta_+$ is just
\begin{equation}
S(\theta_-,\theta_+)=\int_{\theta_-}^{\theta_+}p(t)\>\dot{q}(t)\>\>\rm d t- 
\case12 \>p_+\cdot q_+.
\label{area}
\end{equation}
Hence, the second derivative of the chord action is simply
\begin{equation}
\frac{\partial^2 S}{\partial\theta_+^2}=\case12\>
(p_+\cdot\ddot{q}_+ - q_+\cdot\ddot{p}_+),
\label{hessian}
\end{equation}
so that, as $\theta_+\rightarrow\theta_-$, we obtain
\begin{equation}
\frac{\partial^2 S}{\partial\theta_+^2}\longrightarrow\case12\>\dot{x}_+\wedge\dot{x}_-.
\label{hessian1}
\end{equation}
Therefore the hessian determinant for the stationary phase along $\theta_+=\theta_-$
is just
\begin{equation}
\left|\det\frac{\partial^2 S}{\partial\theta_+^2}\right|=
\left({1\over 2}\right)^l\>|det\{I_-,I_+\}|.
\label{hessian2}
\end{equation}
As noted previously, the Poisson bracket matrix has $l\times l$ dimensions, but 
the power of its preceding factor in \eref{hessian2} is not the same as in
\eref{Jacobian1}, so that finally,
\begin{equation}
\tr\hat{\rho}\approx\>\left({\sqrt2\over2}\right)^l.
\label{trrho1}
\end{equation}

The factor previously  obtained using noncanonical coordinates on the torus,
in \cite{Berry1} for $l=1$, was $\sqrt{2/3}$, instead of \eref{trrho1}.
The important point is not the precise factor, but the fact that this depends neither
on Planck's constant, nor on the classical system, except for the number of degrees of
freedom. There is no point in trying to divide the semiclassical Wigner function 
by the normalization factor \eref{trrho1}, because this will be inaccurate locally, 
inside the torus, as well as destroying the indirect normalization \eref{rhosq}. The
failure of the normalization of the simple semiclassical Wigner function is a consequence
of its (integrable) singularity precisely in the neighbourhood of the classical manifold, 
which is the region that 
dominates the normalization integral \eref{trrho}. In contrast, the integral for
$\tr\hat{\rho}^2$ depends on the entire oscillatory domain of the Wigner function, so that
indirect semiclassical normalization is exact.

What is the effect of evolution on the normalization? Evidently, both \eref{rhosq}
and \eref{trrho1} are stable with respect to unitary evolution. Because the normalization 
integral is dominated  by the region of small chords, 
the decoherence distance functional
\eref{functional} cancels for the stationary phase condition, $\theta_+=\theta_-$, in 
\eref{trrho}, so that there is no effect of decoherence on the direct normalization.
The evolving semiclassical Wigner function is given correctly by \eref{gevol1} 
while the semiclassical approximation remains valid, i. e. until the amplitude
becomes so concentrated on the classical manifold that the assumption of its smoothness
is no loner valid. It follows that the remarkably simple expression
\begin{equation}
\tr\hat{\rho}^2\approx\int\frac{\rm d\theta_-}{(2\pi)^l}\>\frac{\rm d\theta_+}{(2\pi)^l}
\>\> \exp{-\{D_t[\theta_+,\theta_-]\}^2},
\label{entropy}
\end{equation}
in terms of the decoherence distance functional \eref{functional} over the pair of orbits
that end up on $\theta_+$ and $\theta_-$, is valid within the
present semiclassical theory.


\end{document}